\documentclass[a4paper,11pt]{article}
\pdfoutput=1

\usepackage{jinstpub} 
\usepackage{slashbox} 
\graphicspath{{fig/}}
\usepackage{siunitx}

\title{The impact of neutral impurity concentration on charge drift mobility in n-type germanium}

\author[a] {H. Mei,}
\author[a] {G.-J. Wang}
\author[a] {and G. Yang}
\author[a,b,1]{and D.-M. Mei,\note{Corresponding author.}}

\affiliation[a]{Department of Physics, The University of South Dakota,\\414 E. Clark Street, Vermillion, South Dakota 57069, USA}
\affiliation[b]{School of Physics and Optoelectronic, Yangtze University, \\ 1 Nanhuan Street, Jingzhou 434023, China}

\emailAdd{Dongming.Mei@usd.edu}

\abstract{The impact of neutral impurity scattering of electrons on the charge drift mobility in high purity n-type germanium crystals at 77 Kelvin is investigated. We calculated the contributions from ionized impurity scattering, lattice scattering, and neutral impurity scattering to the total charge drift mobility using theoretical models. The experimental data such as charge carrier concentration, mobility and resistivity are measured by Hall Effect system at 77 Kelvin. The neutral impurity concentration is derived from the Matthiessen's rule using the measured Hall mobility and ionized impurity concentration. The radial distribution of the neutral impurity concentration in the self-grown crystals is determined. Consequently, we demonstrated that neutral impurity scattering is a significant contribution to the charge drift mobility, which has a dependence on the concentration of neutral impurities in high purity n-type germanium crystal.}

\keywords{Charge drift mobility; High-purity Germanium; Neutral impurity }

\arxivnumber{} 

\begin{document}
\maketitle 
\flushbottom

\section{Introduction}
\label{sec:intro}
We have reported that the neutral impurity scattering of holes has impact on the charge drift mobility in high purity p-type germanium at 77 Kelvin~\cite{hm}. We have found that the total charge drift mobility in our p-type germanium crystals is dominated by both neutral impurity scattering and acoustic phonon scattering, and the neutral impurity concentration plays an important role on the charge drift mobility. In this paper, we report the impact of neutral impurity scattering of electrons on the charge drift mobility in high purity n-type germanium crystals at 77 Kelvin. Though n-type germanium detectors are less common than p-type detectors, n-type detectors have several unique features. N-type coaxial HPGe detectors usually use an implantation of boron to make an outer p-contact where p-type detectors commonly use lithium diffusion to make an outer n-contact. The implanted p-contact on the outside forms a dead layer that can be less than 0.3 $\mu$m thick which is much thinner than the n-contact~\cite{rbpm}, which is usually more than 1 mm. This very thin dead layer allows photons of energy down to 3 keV to access to the active region. Therefore, n-type detectors provide a much lower detection threshold compared to p-type detectors. Also, the thin dead layer allows for easier segmentation of the outer contact and the segmentation can have a higher efficiency of background rejection, which makes it an interesting alternative in germanium neutrinoless double-beta decay experiments~\cite{maja}. Finally, n-type high-purity germanium detectors are more resistant to neutron damage~\cite{RBWP}.  

The segmented n-type germanium detectors are extensively used in gamma-tracking experiments~\cite{agata, gretina} and can also be used in future neutrinoless double-beta decay experiments~\cite{lev}.
The charge drift mobility plays an important role in understanding the rise time of charge pulses in gamma-tracking experiments~\cite{agata, gretina} and the charge pulse shape in germanium-based neutrinoless double-beta decay experiments~\cite{gerda, majorana, cooper, david, martin}. After the investigation of p-type, it is necessary to investigate the impact of neutral impurity on the charge drift mobility in n-type germanium crystals. P-type germanium crystals and n-type germanium crystals have different charge carriers with different effective masses, which will make the calculation and measurements different. In this work, the calculation of the charge drift mobility due to different scattering processes in n-type germanium crystals, and the impact of neutral impurity concentrations on the charge drift mobility are presented in Section~\ref{s:mobi}, followed by the experimental results on n-type germanium crystals in Section~\ref{s:exp}. Finally, we summarize our conclusions in Section~\ref{s:conc}.

\section{Calculation of Charge Drift Mobility}
\label{s:mobi}
Similar to the calculation we presented in the paper about p-type germanium crystal~\cite{hm}, the total charge drift mobility is impacted independently by different scattering mechanisms and can be determined using Matthiessen's rule~\cite{da}:
\begin{equation}
\frac{1}{\mu_{T}} = \frac{1}{\mu_{I}}+\frac{1}{\mu_{N}}+\frac{1}{\mu_{A}}+\frac{1}{\mu_{O}}+\frac{1}{\mu_{D}},
  \label{e:mu}
\end{equation}
where $\mu_{T}$ is the total charge drift mobility, $\mu_{I}$, $\mu_{N}$, $\mu_{A}$, $\mu_{O}$ and $\mu_{D}$ are the contributions to the mobility from scattering of ionized impurities, neutral impurities, acoustic phonons, optical phonons and dislocations, respectively. For n-type germanium, the majority charge carriers are electrons. The effective mass $m^{\ast}$ of electrons is taken as $m_{e,l}^{\ast }$= 1.64$m_{0}$ for longitudinal direction and $m_{e,t}^{\ast }$= 0.082$m_{0}$~\cite{sinj} for transverse directions, respectively, where $m_{0}$ is the mass of the electron in a vacuum. Then the harmonic mean is used to calculate the conductive effective mass of electrons~\cite{kmsu} :
\begin{equation}
m_{e}^{\ast }=(\frac{3}{\frac{1}{m_{l}}+\frac{1}{m_{t}}+\frac{1}{m_{t}}})m_{0}
\end{equation}
which leads $m_{e}^{\ast }$= 0.12$m_{0}$.

As indicated by eq.~\ref{e:mu}, the mobility due to neutral impurity scattering, $\mu_{N}$, can be evaluated if $\mu_{T}$, $\mu_{I}$, $\mu_{A}$, $\mu_{O}$ and $\mu_{D}$ are known. Furthermore, the neutral impurity concentration, $N_{n}$, can be estimated if the relationship between $N_{n}$ and $\mu_{N}$ can be established.

\subsection{Ionized impurity scattering}
The mobility due to ionized impurity scattering $\mu_{I}$, can be calculated by the CW model~\cite{ec}. It can be simplified for germanium at $T$ = 77 Kelvin as:
\begin{equation}
\mu _{I}=\frac{1.65\times 10^{21}}{N_{i}}\left [ ln(1+\frac{4.01\times 10^{12}}{N_{i}^{2/3}}) \right ]^{-1}
\end{equation} 

A more accurate model developed by Brooks and Herring ~\cite{bh} can be simplified as follows when $T$ = 77 Kelvin:
\begin{equation}
\mu _{I}=\frac{1.65\times 10^{21}}{N_{i}}\left [ ln\frac{1.48\times 10^{18}}{N_{i}} \right ]^{-1}
\label{e:mui2}
\end{equation}
where $\mu_{I}$ is in cm$^{2}$/(V$\cdot$s) and $N_{i}$ is in /cm$^3$. The details for these two models have been discussed from Eq.2.4 to Eq.2.7 in our previous work~\cite{hm}.

\begin{figure}
\centering
\includegraphics[angle=0,width=12.cm] {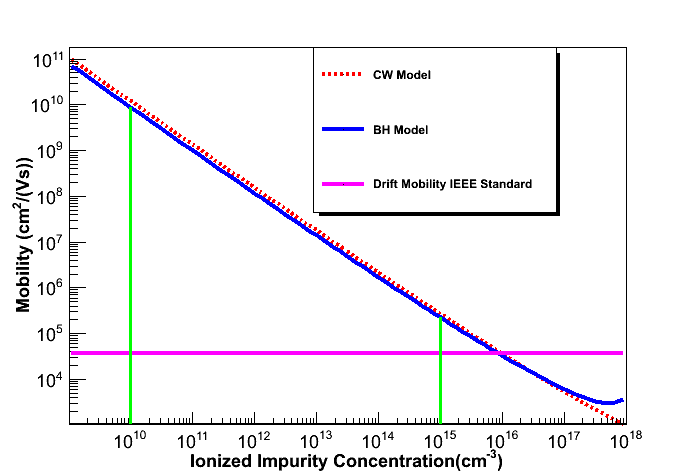}
\caption{The mobility contributed by ionized impurity scattering ($\mu_{I}$) as a function of the ionized impurity concentration. Note that when the ionized impurity concentration is in the region between 10$^{10}$/cm$^3$ and 10$^{15}$/cm$^3$, the total charge drift mobility contributed by the ionized impurity scattering is very small.   }
\label{fig:ionized}
\end{figure}

Eq.~\ref{e:mui2} indicates that $\mu_{I}$ decreases as $N_{i}$ increases at a given temperature. Fig.~\ref{fig:ionized} shows the relationship between the mobility due to ionized impurity scattering ($\mu_{I}$) and ionized impurity concentration. Based on the IEEE Standard~\cite{IEEE}, the value of the electron drift mobility is $\mu_{n}$=36000 cm$^2$/(V$\cdot$s) and $\mu_{p}$=42000 cm$^2$/(V$\cdot$s) for n-type and p-type high-purity germanium crystals, respectively. As shown in Fig.~\ref{fig:ionized}, similar to the results we obtained in the p-type work, it is clear that the total charge drift mobility is governed by the ionized impurities only when the ionized impurity concentration is higher than $\sim$10$^{16}$/cm$^{3}$. The ionized impurity concentration in detector-grade germanium crystals must be in the order of a few times 10$^{10}$/cm$^3$~\cite{EEHA}. With this very low ionized impurity concentration, the mobility due to ionized impurity scattering is of the order of 10$^{9}$ cm$^{2}$/(V$\cdot$s) as indicated in Fig.~\ref{fig:ionized}. Thus, the contribution from the ionized impurity scattering to the total charge drift mobility is very small in general.

\subsection{Acoustic phonon scattering}
The mobility due to acoustic deformation potential scattering, $\mu_{A}$, can be calculated by Eq.2.12~\cite{hm} and can be simplified as
\begin{equation}
{\mu}_{A}=\frac{4.65\times 10^{5}}{m^{\ast 5/2}}\cdot T^{-3/2},
\end{equation}
where $\mu_{A}$ is in cm$^{2}$/(V$\cdot$s). Taking $m_{e}^{\ast }$= 0.12$m_{0}$, we can get that for n-type germanium, the electron drift mobility caused by acoustic phonon scattering is 
\begin{equation}
\mu_{A}=9.32\times 10^{7}\cdot T^{-3/2},
\label{e:mua2}
\end{equation}

\begin{figure}
\centering
\includegraphics[angle=0,width=12.cm] {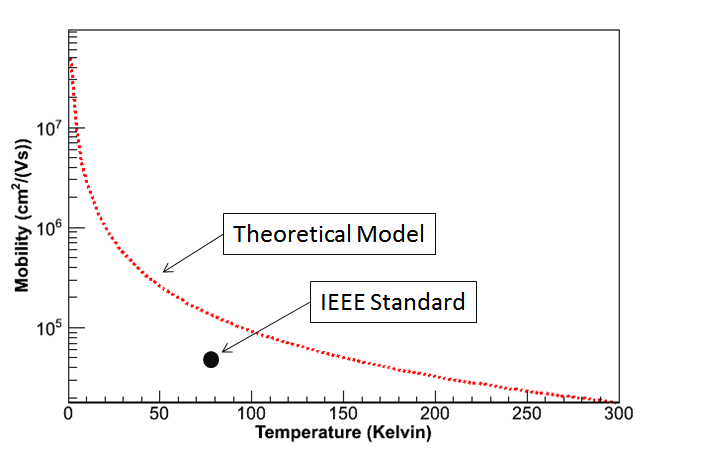}
\caption{The mobility due to acoustic phonon scattering ($\mu_{A}$) as a function of  temperature. The black dot indicate the value of IEEE Standard at 77 Kelvin.}
\label{fig:acoustic}
\end{figure}

Eq.~\ref{e:mua2} implies that there is a temperature dependence in $\mu_{A}$. With $T$ = 77 Kelvin, one obtains $\mu_{A}$ = 1.38$\times$10$^{5}$cm$^{2}$/(V$\cdot$s). This value is very close to our results for p-type germanium crystals where $\mu_{A}$ = 1.15$\times$10$^{5}$cm$^{2}$/(V$\cdot$s). The discrepancy is due to the effective mass of electrons are smaller then the effective mass of holes, which results in a slightly larger value. Fig.~\ref{fig:acoustic} shows the variation of $\mu_{A}$ with temperature, it is clear that the theoretical results of $\mu_{A}$ are larger than the IEEE Standard at 77 Kelvin which indicates that acoustic phonon scattering may not be the sole scattering source of total charge drift mobility at 77K, there must be other scattering mechanisms affecting to the total charge drift mobility.

\subsection{Neutral impurity scattering}
The derivation and modification for the equations about the calculation for neutral impurity scattering has been well described in our previous work~\cite{hm}. With all constants replaced by their corresponding values, electron mobility due to neutral impurity scattering can be simplified as:

\begin{equation}
\mu_N=(\mu _{N})_E=\frac{1.07\times 10^{20}}{N_{n}}
\label{e:neu100}
\end{equation}

\begin{figure}
\centering
\includegraphics[angle=0,width=12.cm] {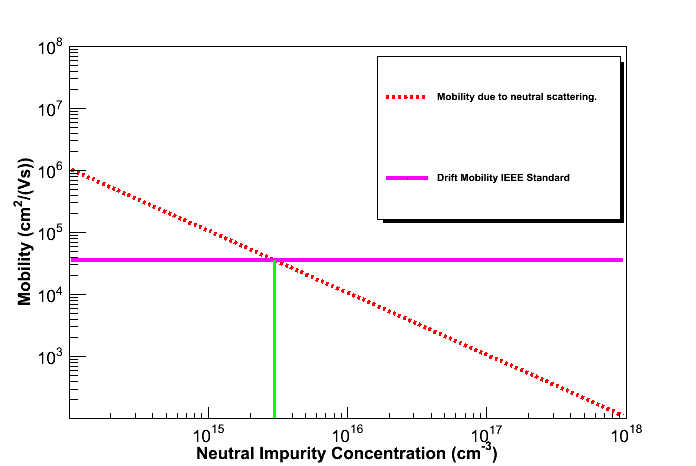}
\caption{The mobility due to neutral impurity scattering calculated by Erginsoy's model ($\mu_{N}$) as a function of  neutral impurity concentration. }
\label{fig:neu1}
\end{figure}

As shown in Fig.~\ref{fig:neu1}, the mobility due to neutral impurity scattering decreases as the neutral impurity concentration increases. With the neutral impurity concentration in the level between 10$^{14}$/cm$^3$-10$^{16}$/cm$^3$, the mobility due to neutral impurity scattering could be an important contribution to the total charge drift mobility at a level of close to the IEEE standard stated earlier.
From Erginsoy's work (eq.~\ref{e:neu100}), we can see that $\mu_{N}$ is temperature independent. We still need to consider the temperature dependent regime. In Sclar's work~\cite{ns, tcm}, a weak dependence of $\mu_N$ on the temperature for semiconductors is:
\begin{equation}
\mu_N=(\mu _{N})_S=0.82(\mu _{N})_E[\frac{2}{3}(\frac{k_{B}T}{E_{N}})^{1/2}+\frac{1}{3}(\frac{E_{N}}{k_{B}T})^{1/2}],
\label{e:neu111}
\end{equation}
where $(\mu _{N})_E$ is the temperature-independent mobility given by eq.~\ref{e:neu100} and $E_{N}$ is the scaled binding energy for the negative ion, $E_{N}$=0.71eV $m^{\ast }$/$m_{e}$($\varepsilon_{r}\varepsilon_0)^2$. 

If we assume the neutral impurity concentration is 2$\times$10$^{15}$/cm$^3$ as measured by \cite{eeh}, then eq.~\ref{e:neu100} becomes a constant, 5.35$\times$10$^{4}$ cm$^{2}$/(V$\cdot$s), and eq.~\ref{e:neu111} becomes: 
\begin{equation}
\mu_N=(\mu _{N})_S=5.35\times 10^{4}(0.28T^{1/2}+0.54T^{-1/2}),
\label{e:neu2}
\end{equation}
which yields that $(\mu _{N})_S$= 1.35$\times$10$^{5}$ cm$^{2}$/(V$\cdot$s) at 77 Kelvin.

\begin{figure}
\centering
\includegraphics[angle=0,width=12.cm] {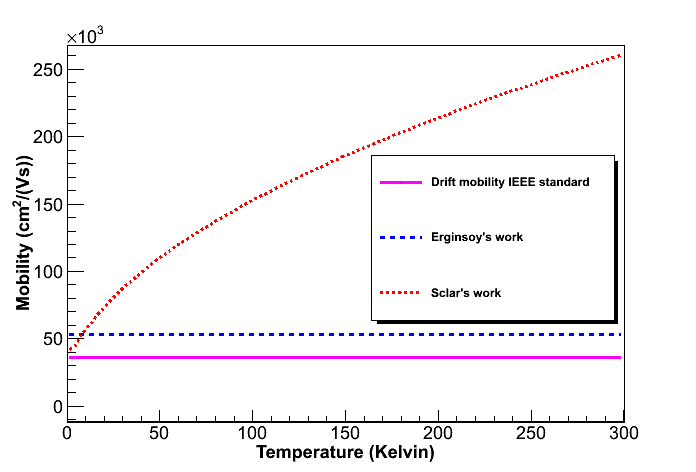}
\caption{The comparison among the work by Erginsoy, Sclar, and IEEE standard for mobility due to neutral impurity scattering with the assumption that the neutral impurity concentration is 2$\times$10$^{15}$/cm$^3$.}
\label{fig:neutral-com}
\end{figure}

\begin{figure}
\centering
\includegraphics[angle=0,width=12.cm] {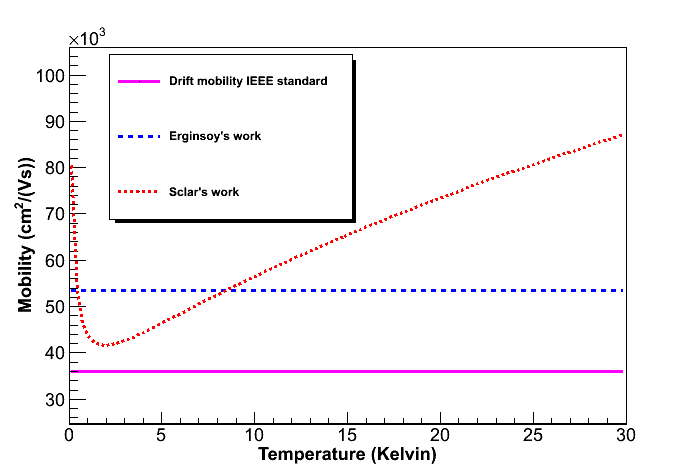}
\caption{The calculated freeze-out temperature of our germanium crystals is around 2 Kelvin.}
\label{fig:neutral-com.1}
\end{figure}

As shown in Fig.~\ref{fig:neutral-com}, there is a factor of two difference for the mobility due to neutral impurity scattering among the work by Erginsoy and Sclar at 77 Kelvin. For the same reason, we continue following  Erginsoy's theory in this work, i.e. $\mu_{N}$ has no temperature dependence as the freeze-out temperature of our germanium crystals is around 2 Kelvin shown in Fig.~\ref{fig:neutral-com.1}.

\subsection{Other scatterings}
For the mobility caused by the scattering of optical phonons in germanium, $\mu_{O}$, it can be ignored for the same reason as we discussed in our previous work~\cite{hm,bs,ssl}. The mobility due to dislocation scattering, $\mu_{D}$ can also be ignored since the dislocation density for n-type high purity germanium crystal is similar to that of p-type high purity germanium crystal ~\cite{hm,dt,gwy}.

\section{Experimental Results}
\label{s:exp}
Several germanium samples, obtained from a n-type detector-grade crystal grown in our lab at the University of South Dakota~\cite{guo1, gang1, gang2, guo3, gang3, gang4, gang5, guo4} with measured average Hall mobility $\mu_{H}$ larger than 30000 cm$^2$/(V$\cdot$s), are used for our investigation in this work. Based on IEEE Standard, the relationship between the measured Hall mobility $\mu_{H}$ and the total charge drift mobility $\mu$ is defined as:
\begin{equation}
\mu=\frac{\mu _{H}}{r},
\label{e:mun4}
\end{equation}
where $r$ is a constant near unity depending on crystal orientation and magnetic field B. In our Hall Effect measurements, we use magnetic field B=0.5 T. The factor $r$ is assumed to be 1.03 for p-type high-purity germanium crystals in our previous work. For n-type high-purity germanium crystals, the factor $r$ is assumed to be 0.93 when B= 0.5 T~\cite{IEEE}. This means the actual charge drift mobility would be larger than the measured Hall mobility for the n-type germanium crystals.

\begin{figure}
\centering
\includegraphics[angle=0,width=10.cm] {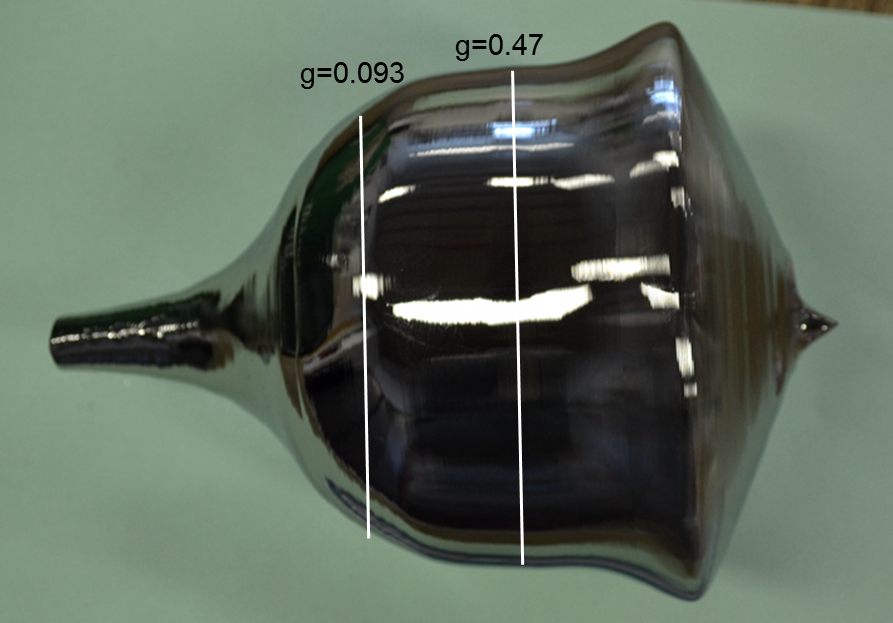}
\caption{Two wafers are cut from the detector grade n-type high purity germanium crystal with the white lines indicating the position. The parameter, $g$,  denotes the fraction of original liquid which is frozen. }
\label{fig:cut1}
\end{figure}

\begin{figure}
\centering
\includegraphics[angle=0,width=10.cm] {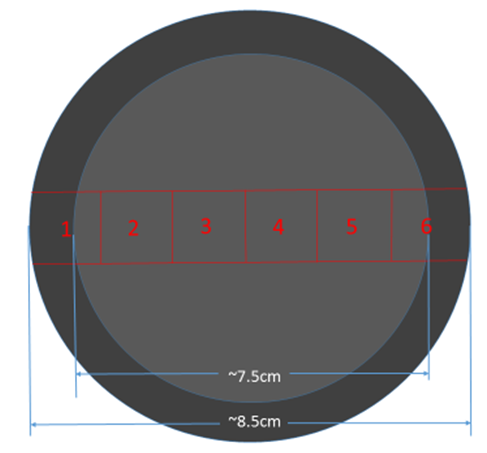}
\caption{Schematic diagram showing the location of the samples cut from the three wafers.}
\label{fig:cut2}
\end{figure}

Two wafers are cut from the detector-grade crystal mentioned above at different axial positions in Fig.~\ref{fig:cut1}. The symbol $g$, is the fraction of the melt that has been crystallized. Five square samples are cut from the wafer with g=0.093, and six samples are cut from the wafer with g=0.47 as shown in Fig.~\ref{fig:cut2}. These samples are prepared in a manner similar to that used for p-type germanium crystals prior to using the Van der Pauw Hall Effect Measurement System to measure their electrical properties at 77 Kelvin. The only difference is we use Sn-Sb(Sn:Sb=95:5) to make ohmic contacts for n-type samples instead of using In-Ga for p-type samples. The same uncertainty(5\%) for the Hall effect measurements is applied~\cite{kab}.

\begin{figure}
\centering
\includegraphics[angle=0,width=12.cm] {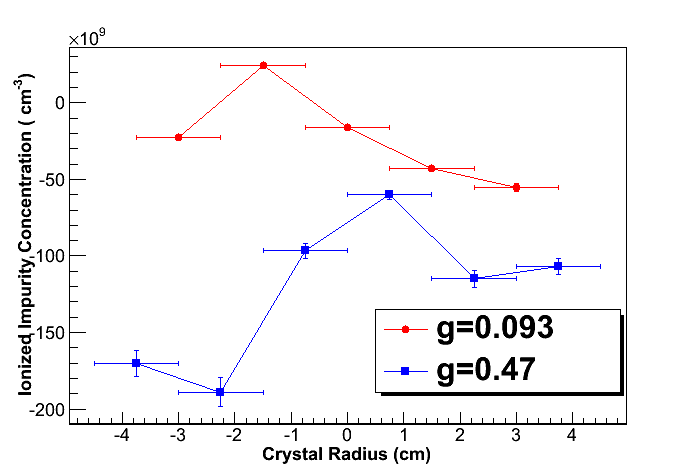}
\caption{The net carrier concentration as a function of crystal radius for all germanium samples.}
\label{fig:concentration}
\end{figure} 

\begin{figure}
\centering
\includegraphics[angle=0,width=12.cm] {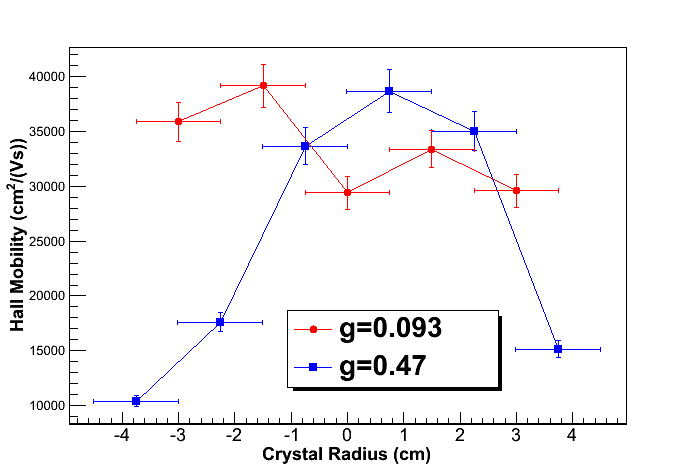}
\caption{The radial distribution of Hall mobility of the three wafers.}
\label{fig:mobility}
\end{figure}

\begin{figure}
\centering
\includegraphics[angle=0,width=12.cm] {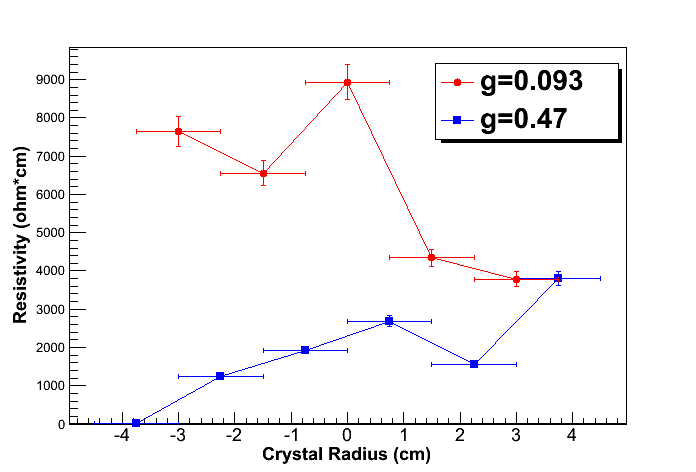}
\caption{The radial distribution of resistivity of the three wafers.}
\label{fig:resistivity}
\end{figure}

Fig.~\ref{fig:concentration} shows the ionized impurity concentration as a function of crystal radius for all germanium samples. As indicated by Fig.~\ref{fig:concentration} and similar to the p-type germanium crystal, the lower ionized impurity concentration is observed at the center part of the crystal. For sample 2 on slice 1, it is the only sample with positive ionized impurity concentration. This makes the electrical properties for samples 1, 2 and 3 on slice 1 complicated due to the existence of a p-n junction. This also indicates a nonuniform impurity distribution along the axial direction. Fig.~\ref{fig:mobility} and Fig.~\ref{fig:resistivity} show the radial distribution of Hall mobility and resistivity for all samples, respectively. 

\begin{figure}[htb!]
\centering
\includegraphics[angle=0,width=12.cm] {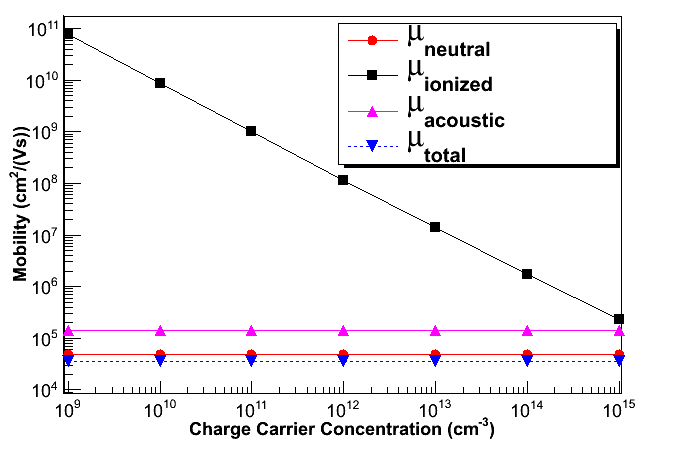}
\caption{The relationship between the charge carrier concentration and all the scattering processes at 77 Kelvin.}
\label{fig:den}
\end{figure}

According to our Hall Effect measurements shown in Fig.~\ref{fig:mobility}, our germanium crystals have a Hall mobility $\mu_{T}$ of $\sim$ 30000 cm$^2$/(V$\cdot$s) which leads the total drift mobility to $\sim$ 36000 cm$^2$/(V$\cdot$s) meeting the IEEE Standard. The measured ionized impurity concentration is in the range of -1.61$\times$10$^{10}$/cm$^3$ to -1.83$\times$10$^{11}$/cm$^3$ as shown in Fig.~\ref{fig:concentration}. Using eq.~\ref{e:mui2}, the calculated mobility due to this level of the ionized impurity scattering $\mu_{I}$ is in the range of 5.76$\times$10$^{8}$ to 5.59$\times$10$^{9}$ cm$^{2}$/(V$\cdot$s). Similarly, from eq.~\ref{e:mua2}, when $T$ = 77 Kelvin, the mobility due to acoustic phonon scattering $\mu_{A}$ = 1.38$\times$10$^{5}$cm$^{2}$/(V$\cdot$s). With $\mu_{O}$ as well as $\mu_{D}$ being ignored, the mobility due to neutral impurity scattering $\mu_{N}$ can be deduced using eq.~\ref{e:mu}. Our deduced results showed that $\mu_{N}$ is almost a constant, 4.87$\times$ 10$^{4}$cm$^{2}$/(V$\cdot$s) when the ionized impurity concentration is in the range of 10$^{10}$/cm$^3$ to 10$^{15}$/cm$^3$ as shown in Fig.~\ref{fig:den}. Since the deduced $\mu_{N}$ and the calculated $\mu_{A}$ are much smaller than $\mu_{I}$ for a detector-grade crystal at 77 Kelvin, we conclude that the total charge drift mobility ($\mu_{T}$) in our germanium crystals is dominated by both $\mu_{N}$ and $\mu_{A}$, and the neutral impurity concentration has important impact on the charge drift mobility.

\begin{figure}[htb!]
\centering
\includegraphics[angle=0,width=12.cm] {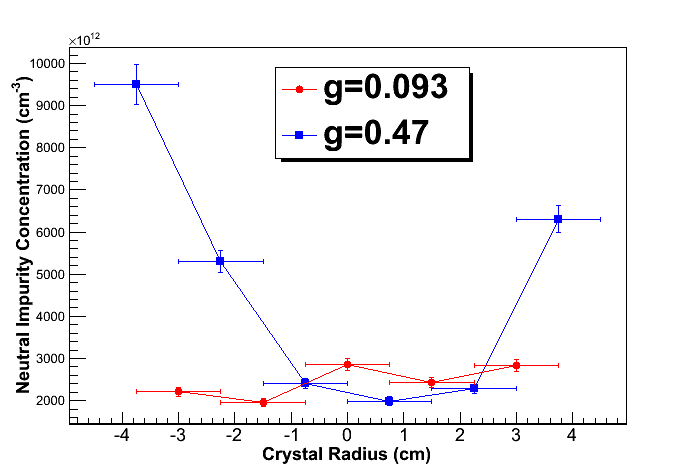}
\caption{The radial distribution of neutral impurity concentration.}
\label{fig:net}
\end{figure}

Once $\mu_{N}$ is obtained, the neutral impurity concentration, $N_n$, can then be calculated from eq.~\ref{e:neu100}, which yields that $N_n$ is in the range of 1.95$\times$10$^{15}$/cm$^3$ to 9.5$\times$10$^{15}$/cm$^3$ when the ionized impurity concentration is in the range of 10$^{10}$/cm$^3$ to 10$^{15}$/cm$^3$. Especially when the total charge drift mobility is $\sim$ 36000 cm$^2$/(V$\cdot$s), the calculated neutral impurity concentration is 2.2$\times$10$^{15}$/cm$^3$.
This level of neutral impurity concentration agrees with the pioneering work (2$\times$10$^{15}$/cm$^{3}$) by Hansen et al. in 1982~\cite{wlh}.
Fig.~\ref{fig:net} shows the relationship between the calculated neutral impurity concentration ($N_n$) and the crystal radius. Fig.~\ref{fig:net} implies that there are more neutral impurities at the edge than in the center of the crystal. This is very similar to the case of the radial distribution of ionized impurities, where the neutral impurity is different, the total charge drift mobility can be different. 

The source for neutral impurities would be the same as was determined from our previous analysis. We may have carbon, silicon, hydrogen, nitrogen and oxygen introduced during zone-refining; hydrogen, silicon and oxygen introduced during crystal growth. Neutral impurities are electric inactive atoms, and can not be measure by the Hall Effect system. No matter n-type or p-type Ge crystals, even though the ionized impurities concentration are quite different, the neutral impurity concentration should be in the similar level. Based on our theoretical model and experimental results, our results shows that the neutral impurity concentration is 1.95$\times$10$^{15}$/cm$^3$ to 9.5$\times$10$^{15}$/cm$^3$ for n-type Ge crystals and 2.8$\times$10$^{15}$/cm$^3$ to 5$\times$10$^{15}$/cm$^3$ for p-type Ge crystals~\cite{hm}, which are of very similar levels. And the discrepancy can be explained by the different effective mass for the charge carriers which results in a different calculation; or because of the different ohmic contact materials applied before the Hall Effect measurements.

\section{Conclusions}
\label{s:conc}
The conclusion is similar to the work we did previously~\cite{hm}. We investigate the scattering mechanisms that contribute to the total charge drift mobility in n-type detector-grade germanium crystals. We evaluate the neutral impurity concentration from measured Hall Effect results and theoretical model. We found that for high-purity germanium crystal along $<100>$ direction, with impurity level of 10$^{10}$/cm$^3$-10$^{11}$/cm$^3$ and dislocation density below 10$^{4}$/cm$^2$, the neutral impurity scattering is an important scattering mechanism at 77 Kelvin. Very similar to the results in p-type Ge crystal, there are more neutral and ionized impurity atoms at the edge part of the crystal than that at the center part. This results in the lower charge drift mobility at the edge part and higher charge drift mobility at the center part. 

\section*{Acknowledgments}
The authors wish to thank Christina Keller for her careful reading of this manuscript. We also would like to thank Jing Liu for his useful discussion.  This work is supported in part by NSF PHY-0919278, NSF PHY-1242640, NSF OIA 1434142, DOE grant DE-FG02-10ER46709, the Office of Research at the University of South Dakota and a research center supported by the State of South Dakota.






\end{document}